\documentclass[3p,amsmath,amssymb,compress]{elsarticle}
\usepackage{amsmath}
\usepackage{graphicx}
\usepackage{epsfig}
\usepackage{latexsym}
\usepackage{amsmath,amssymb}
\usepackage{bm} 
\usepackage{xcolor}
\usepackage{stackrel}
\usepackage{subcaption}
\usepackage{accents}
\usepackage{ulem}
\usepackage{bbm}
\usepackage{comment}
\def\thefootnote{*}\footnotetext{J.K. and E.P. contributed equally to this work.}

\journal{Annals of Physics special issue dedicated to the memory of Konstantin Efetov}

\numberwithin{equation}{section}

\DeclareMathOperator{\tr}{tr}
\DeclareMathOperator{\Tr}{Tr}
\def\d{\mathrm{d}}

\newcommand{\tgl}{\tau_{\mathrm{GL}}}
\newcommand{\trel}{\tau_{\mathrm{R}}}
\newcommand{\tc}{T_c}

\begin{document}
\begin{frontmatter}	


\title{Fluctuation-driven excess noise near superconducting phase transition}

\author[UW]{Juhun Kwak\thefootnote{}}
\author[UW]{Emil Pellett\thefootnote{}}
\author[SSR]{Elio J. K\"onig}
\author[UW]{Alex Levchenko}

\address[UW]{Department of Physics, University of Wisconsin-Madison, Madison, Wisconsin 53706, USA}
\address[SSR]{Max-Planck-Institut f\"ur Festk\"orperforschung, 70569 Stuttgart, Germany}

\date{December 28, 2022}

\begin{abstract}
We discuss intrinsic mechanisms of nonequilibrium excess noise in superconducting devices and transition edge sensors. In particular, we present an overview of fluctuation-driven contributions to the current noise in the vicinity of the superconducting transition. We argue that sufficiently close to the critical temperature fluctuations of conductivity may become correlated provided that the rate of quasiparticle relaxation is slow as compared to dynamics of superconducting fluctuations. In this regime, fluctuations of conductivity adiabatically follow the fluctuations of the electron distribution. This leads to a substantial enhancement of current noise. The corresponding spectral power density of noise has a Lorentzian shape in the frequency domain while its magnitude scales proportionally to the inelastic relaxation time. It also sensitively depends on the dephasing and Ginzburg-Landau timescales. Further estimates suggest that this mechanism dominates over the conventional temperature fluctuations in the same range of parameters. To describe these effects microscopically, we use the nonequilibrium Keldysh technique in the semiclassical approximation of superconductivity with Boltzmann-Langevin random forces to account for correlations of fluctuations.     
 \end{abstract}

\begin{keyword}
Shot noise, Fano factor, superconducting fluctuations, Boltzmann-Langevin theory, Keldysh formalism 
\end{keyword}
\end{frontmatter}
\newpage

\section{Introduction}  

The fluctuation-dissipation theorem (FDT) establishes a direct relation between the equilibrium Johnson-Nyquist (JN) noise power density of electric current fluctuations $S_0=4TG$
and the linear differential conductance $G$ in the system at finite temperature $T$ \cite{Kogan}. In this sense, measuring the equilibrium current noise provides the same information 
about electric transport as the linear response measurement. The FDT does not hold for nonequilibrium processes. In general, noise contains much richer information 
about quasiparticle kinetics. The frequency dependence of the noise allows for the determination of the timescales of inelastic relaxation. The shot noise can reveal
the fundamental properties of the charge carriers. For example, the fractional charge of Laughlin quasiparticles $e^*=e/3$ was determined from the shot noise measured in the fractional quantum Hall effect \cite{Reznikov}. 

In this paper, we describe the intrinsic excess noise mechanisms in superconducting devices. Apart from the conceptual interest, this work is also motivated by multiple experimental results on unconventional superconductors \cite{LeeLeePRB89}, microconstrictions \cite{Reznikov2007}, and particularly on thin films \cite{KhrapaiArxiv1,KhrapaiArxiv2} that reported enhanced noise observed across the resistive transition. 
In terms of practical device applications the study of excess noise continues to attract considerable interest in the context of transition edge sensors (TES) \cite{IrwingHilton,WesselsAPL21,GottardiPRL21}.     
TES are superconducting devices operated in the regime close to the resistive transition and employed, for example, for particle detection in astrophysics applications. Given their extreme sensitivity, TES are effective thermometers and bolometers over a wide range of wavelengths. At the same time, due to this very sensitivity, a thorough understanding of noise mechanisms is of imperative relevance.

\subsection{Noise in mesoscopic conductors}

The crossover from equilibrium to shot noise can distinguish between the regimes of mesoscopic transport (see review Ref. \cite{ButtikerBlanter} and references therein). In particular, the zero-frequency noise in a disordered metallic wire depends on a relationship between the wire length and the electron mean free path. In relatively short wires, electrons diffuse without experiencing inelastic collisions. 
In this limit, the nonequilibrium distribution of electrons in the wire is given by the linear superposition of electron distributions in the leads. This results in a noise spectrum of the form \cite{NagaevPLA92,ALY,SukhorukovPRL98}
\begin{equation}\label{eq:noise-N1}
\frac{S}{S_0}=\frac{2}{3}+\frac{v}{6}\coth\left(\frac{v}{2}\right),
\end{equation}
where $v=eV/T$ is the applied voltage normalized to temperature. In the shot noise regime, $eV\gg T$, one finds $S=2eFGV$ with the Fano factor $F=1/3$,
which distinguished noise in a wire as compared to that of the Schottky noise through a tunnel barrier. In longer wires, electron-electron collisions may become more efficient and lead to local equilibration. 
In this case, the nonequilibrium electronic distribution can be found in the form of the usual Fermi-Dirac function but with the spatially varying voltage-dependent temperature. This gives the following expression for the noise power \cite{KozubRudin,NagaevPRB95,deJongPA96}    
\begin{equation}\label{eq:noise-N2}
\frac{S}{S_0}=\frac{1}{2}+\frac{v}{2}\left(\frac{2\pi}{\sqrt{3}v^2}+\frac{\sqrt{3}}{2\pi}\right)
\arctan\left(\frac{\sqrt{3}v}{2\pi}\right),
\end{equation}
and thus a different Fano factor $F=\sqrt{3}/4$. Even though the numerical difference between the Fano factors in these two limiting cases of cold and hot electrons is small, it was shown that electron collisions impact strongly higher cummulants of current fluctuations \cite{BagretsPRL04,GutmanPRB05}.  

\subsection{Noise in mesoscopic superconductors}

In the presence of a superconducting element the current noise in the circuit can be altered substantially. When one of the leads connecting the wire is a superconductor the noise doubles its value \cite{deJongPRB94}. At the subgap voltages, $\{T,eV\}\ll\Delta$, where $\Delta$ is the energy gap, noise is approximately given by Eq. \eqref{eq:noise-N1} with the replacement of the charge $e\to2e$. This doubling is the result of Andreev reflections of quasiparticles at the superconductor-normal (SN) interface. At voltages above the gap, $eV>\Delta$, the shot noise restores its normal state value but acquires an additional voltage-independent excess contribution, which is determined by the gap and the normal state conductance, namely $S=\frac{2}{3}(eV+\Delta)G$ \cite{KhlusJETP87,NagaevPRB01}.  

In addition, superconductivity manifests itself via the proximity effect, that extends over distances on the order of the coherence length, into the wire. This modifies the spectral flow, since subgap quasiparticles have an energy-dependent effective diffusion coefficient. Thus, the proximity effect leads to the so-called reentrant behavior \cite{ArtemenkoSSC79,StoofPRB96,BelzigSM99}: the conductance of the wire first increases as a function of temperature, $G_{\text{SN}}/G=1+a(T/E_{\text{Th}})^2$, at $T< E_{\text{Th}}\ll\Delta $, and then decreases as $G_{\text{SN}}/G=1+b\sqrt{E_{\text{Th}}/T}$ when $E_{\text{Th}}< T\ll\Delta$, where $E_{\text{Th}}$ is the Thouless energy, and $a,b$ are the numerical coefficients. Therefore, the current-voltage characteristic $I=I(V)$ displays the reentrance, which subsequently modifies noise \cite{HouzetPRB03}  
\begin{equation}
S=\frac{8T}{3}\frac{\partial I}{\partial V}+\frac{4eI}{3}\coth\left(\frac{eV}{T}\right).
\end{equation}
This expression captures the excess noise even at the subgap energies $\{T,E_{\text{Th}},eV\}\ll\Delta$. Numerical analysis shows that noise peaks at $eV\sim E_{\text{Th}}$ and exceeds its normal state value by roughly a factor of two. These examples suggest that current noise increases additionally when superconductivity is introduced into the bulk of the sample and not only as a boundary condition.  

\subsection{Noise in superconductors near $T_c$}

A particularly interesting situation arises in superconducting systems near the resistive transition above the critical temperature, $\Delta T=T-T_c\ll T_c$. In this regime, transport properties are strongly affected
by spontaneous generation and dissociation of Cooper pairs. It is perhaps surprising that despite the fact that we learned a lot about superconducting fluctuation corrections to the linear response transport coefficients \cite{LarkinVarlamov,VarlamovRMP18}, the problem of nonequilibrium current noise received very limited attention \cite{NagaevPC91,ShimshoniPRB93,MartinPRB00,ALPRB08,DBAL}. Below we recall key results from Ref. \cite{NagaevPC91} where leading contributions to the noise spectral density were identified. We also expand on the scope of considerations, in particular concerning the limits of applicability of the obtained results. This will help us introduce proper terminologies and place our finding in the context of existing works. 

The analysis of Ref. \cite{NagaevPC91} was based on the diagrammatic computation of the current-current correlation function in a perturbative expansion to the leading order in electric field. One-loop diagrams were calculated in the Keldysh technique \cite{Keldysh65}, thus bypassing complications of analytical continuation present in the more standard Kubo formulas of the Matsubara approach \cite{Mahan}. The identified contributions have their origins in different fluctuation effects, all of which have direct analogs to respective terms in the DC conductivity. Therefore, the existing terminological classification of fluctuation-induced corrections remains appropriate. 

The first contribution to noise comes directly from the fluctuation-induced Cooper pairs accelerated by the field. In DC conductivity, this source corresponds to the regular Aslamazov-Larkin (AL) correction \cite{AL68}. 
For a square film of dimensions $L\times L$ the corresponding derived noise correction is of the form:            
\begin{equation}\label{eq:S-reg}
\delta S_{\text{reg}}(\Omega)\simeq T\sigma_{\text{AL}}XP(\Omega\tgl),\quad X=\frac{E_{\text{Th}}(eV)^2}{T^3_c}(T_c\tgl)^3, 
\end{equation}
In the above expression, $\sigma_{\text{AL}}$ is the AL conductivity. For a two-dimensional film $ \sigma_{\text{AL}}/\sigma_{\text{N}}=(T_c\tgl)/(2\pi g)$, where $g$ is the dimensionless sheet conductance, $\sigma_{\text{N}}$ is the normal state conductivity, and $\tgl=(\pi/8)(T-\tc)^{-1}$ is the Ginzburg-Landau time. The Thouless energy has a standard definition $E_{\text{Th}}=D/L^2$ with $D$ being the diffusion coefficient. The spectral density of noise vanishes at small frequencies as $P(w)\propto w^2$ for $w=\Omega\tgl\ll1$, and also decays algebraically as $P(w)\propto1/w^2$ at high frequencies $w\gg1$. It should also be anticipated that Eq. \eqref{eq:S-reg} applies only below a certain critical voltage. This is because strong electric field acts as an effective depairing by detuning the fluctuating condensate further away from the zero field critical temperature. To estimate this field we compare the typical energy of the preformed Cooper pair $\sim T-T_c$ to the work done by the field $\sim eE\xi_{\text{GL}}$ on the size of the pair. Near $T_c$ we have $\xi_{\text{GL}}\simeq\sqrt{D/(T-\tc)}$. This gives a critical field strength of the order $E_c\sim(T-T_c)^{3/2}/e\sqrt{D}$. If we now convert this field into the corresponding critical voltage $V_c$ and estimate the parameter $X$ in Eq. \eqref{eq:S-reg} at $V_c$, we see that $X_c\sim1$. Since $X$ serves as a natural expansion parameter in perturbation theory we expect it to break down at $V>V_c$. To extrapolate Eq. \eqref{eq:S-reg} to higher voltages we recall that nonlinear Aslamazov-Larkin conductivity decays as $\sigma_{\text{AL}}(E)\propto (E_c/E)^{2/3}$ for $E>E_c$ \cite{DorseyPRB91}. Since it comes from the same current-current correlation functions diagrammatically, it should be expected that the noise decays similarly, giving $\delta S_{\text{reg}}\propto (V_c/V)^{2/3}$ for $V>V_c$. 

The second contribution to the current noise is connected with the variation of the electron distribution function by the electric field. In the linear response these are the terms responsible for the so-called anomalous Maki-Thompson (MT) conductivity near $T_c$ \cite{MakiPTP68,ThompsonPRB70} accounting for the Andreev reflection from droplets of fluctuating Cooper pairs. Their extension to the current fluctuations yield noise power in the form: 
\begin{equation}\label{eq:S-anom}
\delta S_{\text{anom}}(\Omega)\simeq T\sigma_{\text{AL}}(\tau_\phi/\tgl)XK(\Omega\tgl),\qquad K(w)=\frac{\arctan(w/2)}{w/2}.
\end{equation}
Unlike Eq. \eqref{eq:S-reg}, this term does not vanish in the zero frequency limit. It is worthwhile to stress that the anomalous contribution to the excess noise is more singular than the equilibrium noise associated with the MT conductivity, $\sigma_{\text{MT}}=2\sigma_{\text{AL}}\ln(\tau_\phi/\tgl)$, which diverges only logarithmically with the dephasing time $\tau_\phi$. Due to the anomalous nature of this contribution, it is a more challenging task to extract the noise function asymptote at large voltages. We are able to estimate the suppression of the nonlinear MT conductivity, $\sigma_{\text{MT}}(E)\propto\ln(E_c/E)$, which turns out to be much weaker than that of a regular term. There is also an interference term between the regular and anomalous parts of the current vertices. This contribution to the noise spectral density has the same structural form as Eq. \eqref{eq:S-reg} albeit with a different frequency dependence with a much faster decay at $\Omega\tgl>1$.          

Following the standard terminology, there is a third type of fluctuation-induced correction to the DC conductivity. It is reducing the transport coefficients due to the effective depletion of the electronic density of
states (DoS) as a consequence of electrons virtually forming Cooper pairs \cite{AbrahamsPRB70}. However, this fluctuation effect is subdominant both in the DC conductivity, $\sigma_{\text{DoS}}/\sigma_{\text{N}}\propto -\ln(T_c\tgl)/g$, and noise. 

Another mechanism of noise is the effect of heating in the electric field. The simplest way to estimate noise from this mechanism is via the effective electron temperature approximation, $\delta S_{\text{heat}}\propto 2T(\partial\sigma/\partial T)\delta T$, with $\delta T=D(eE)^2\tau_\text{R}/T$, where $\tau_{\text{R}}$ denotes the quasiparticle energy relaxation time. In 2D films with weak pair-breaking the most significant temperature dependence of conductivity is determined by the MT-term $\sigma_{\text{MT}}(T)$. This gives an estimate for the corresponding noise power in the form  
\begin{equation}\label{eq:S-heat}
\delta S_{\text{heat}}(\Omega)\simeq -T\sigma_{\text{MT}}\frac{E_{\text{Th}}(eV)^2}{T^3_c}(\tc\trel)(\tc\tgl).
\end{equation}
Concerning the spectral properties, it should be remarked that even equilibrium noise above $T_c$ has a complicated frequency dependence, since both $\sigma_{\text{AL}}(\Omega)$ and $\sigma_{\text{MT}}(\Omega)$ AC conductivities behave rather nontrivially \cite{AslamazovJLTP79,PetkovicJPCM13,BurmistrovAoP20}.   

In this work we find another mechanism of excess noise in superconductors near $T_c$. It is expected to be most pronounced in the situations with long electron relaxation and fast superconducting fluctuations when $\tau_{\text{R}}\gg\tau{_\text{GL}}$. Under such nonequilibrium conditions, fluctuations of conductivity $\delta\sigma$ follow adiabatically fluctuations of the electron distribution function $\delta f$, and thus become correlated due to the stochastic nature of electron scattering in the Boltzmann-Langevin description. These fluctuations are distinct from the mesoscopic conductance fluctuations near $T_c$ \cite{ZhouPRL98,HettingerPRB19}.  We determine the corresponding susceptibility $\chi=\delta\sigma/\delta f$ and convert conductivity noise of the resistive transition into the excess current noise of 2D superconducting films where we find a dominant correction of the form (cf. Eq. \eqref{eq:S-I-sigma} derived below) 
\begin{equation}\label{eq:S-corr}
\delta S_{\text{corr}}(\Omega)\simeq T\frac{\sigma^2_{\text{MT}}}{\sigma_{\text{N}}}
(T_c\tau_{\text{R}})XL(\Omega\tau_{\text{R}}),\qquad L(u)=\frac{1}{1+u^2}.
\end{equation}

The remainder of this paper is structured as follows: Sec. \ref{sec:formalism} contains the microscopic derivation of our main result, Eq. \eqref{eq:S-corr}. In this context, we also derive a formula for the fluctuation-induced conductivity for an arbitrary electronic distribution function. In the discussion Sec. \ref{sec:discussion} we compare our result to an estimate of the effect of temperature fluctuations (Sec. 3.1) and conclude with a summary and outlook (Sec. 3.2).


\section{Formalism}\label{sec:formalism}

In this section, we present a microscopic analysis of the fluctuation-correlated mechanism of intrinsic excess noise near the superconducting phase transition. At the technical level, our approach is based on Efetov's nonlinear sigma-model field theory \cite{Efetov} embedded into the Keldysh technique of nonequilibrium superconductivity \cite{SkvortsovPRB00,TDGL,Kopnin}. The calculation consists of three parts. First, we derive expressions for the fluctuation-induced conductivity for an arbitrary electronic distribution. In part this analysis overlaps with earlier works \cite{VolkovPRB98,TikhonovPRB12,SchwietePRB13} that used quasiclassical approximation and the Usadel equation \cite{Usadel,VolkovPRB98,TikhonovPRB12,SchwietePRB13}. We also benchmark obtained expressions against known diagrammatic derivations \cite{LarkinVarlamov}. Next, we employ the Boltzmann-Langevin formalism to establish the correlation function of conductivity fluctuations from the known correlator of the distribution function in the normal state above $T_c$. This approach closely follows earlier ideas of using the Boltzmann-Langevin scheme to determine current noise in the vicinity of the 2D superconductor-insulator quantum critical point \cite{GMSV}. Finally, we provide practical estimates for the strength of calculated noise as compared to other complementary mechanisms described in the introduction.\footnote{Throughout the paper we use dimensionless units $\hbar=k_B=1$ and restore physical units in the final expressions when we make numerical estimates.}

\subsection{Expression for the $Q$-matrix}

Our starting point is the Keldysh nonlinear sigma-model of disordered superconductors \cite{SkvortsovPRB00,TDGL}, see also the review \cite{KNLSM} and the references therein. 
This framework allows us to go beyond the linear response. It also enables a systematic account of the elastic disorder averaging.
The formalism considers the evolution along the closed contour in the time direction. It necessarily deals with the
two replicas of each field such as the vector potential $\bm{A}(\bm{r},t)$ and superconducting order parameter $\Delta(\bm{r},t)$. One of the replicas encodes evolution in
the forward time direction while the other encodes evolution in the backward time direction. The half-sum of these fields is denoted as the classical component; the half-difference is its quantum counterpart. 
The former is the physical field entering the equation of motion. The latter is a counting field used to generate these equations.  
The action of this field theory $\mathcal{A}[Q,\bm{A},\Delta]$ is described by the $4\times4$ matrix-field $\hat{Q}_{tt'}(\bm{r})$, which is also an infinite matrix integral
kernel with respect to its two time indices. The matrix dimension is the direct product of Keldysh and Gor'kov-Nambu subspaces. 
For a short-range correlated disorder the $\hat{Q}$ matrix is a local function of the spatial variable. 
This matrix obeys the local nonlinear constraint $\hat{Q}^2=1$ and technically appears as a Hubbard-Stratonovich (HS) transformation of a four-fermion term after the disorder averaging. 
The $Q$-matrix can be interpreted as the semiclassical Green's function. The field $\Delta(\bm{r},t)$ appears the HS decoupling of the BCS interaction term. 

The variation of the action with respect to the quantum component of the vector potential gives an electrical current $\bm{j}=(1/2)\delta\mathcal{A}/\delta\bm{A}^q$. 
It can be expressed in terms of the $Q$-matrix as follows    
\begin{equation}\label{eq:J}
\bm{j}=\frac{\pi}{2}e\nu
D\iint\frac{\d\varepsilon\d\varepsilon'}{4\pi^2}\tr\left\{\hat{\tau}_{z}\left[\hat{Q}^{R}_{\varepsilon\varepsilon'}(\bm{r})
\partial_{\bm{r}}\hat{Q}^{K}_{\varepsilon'\varepsilon}(\bm{r})+\hat{Q}^{K}_{\varepsilon\varepsilon'}(\bm{r})
\partial_{\bm{r}}\hat{Q}^{A}_{\varepsilon'\varepsilon}(\bm{r})\right]\right\}\,,
\end{equation}
where $\nu$ is a single particle normal state density of states at the Fermi energy. We used the Fourier transform from the time to energy variable $\hat{Q}_{tt'}\to \hat{Q}_{\varepsilon\varepsilon'}$. 
The notation $\hat{Q}^{A/R/K}_{\varepsilon\varepsilon'}$ denotes advanced/retarded/Keldysh components of the $Q$-matrix field, which are still $2\times2$ matrices in the Gor'kov-Nambu subspace. We use a basis of Pauli matrices $\hat{\tau}_{i}$ with $i=(0,x,y,z)$ to work in that space. The symbol $\tr\{\ldots\}$ stands for the matrix trace.

The variation of the action with respect to the $Q$-matrix, $\delta\mathcal{A}/\delta\hat{Q}=0$, defines the saddle-point equation for $\hat{Q}$. In this theory it coincides with the Usadel equation. Its off-diagonal Keldysh block defines the kinetic equation. The diagonal blocks define the causal components that satisfy the following equation  
\begin{equation}\label{eq:Usadel}
D\partial_{\bm{r}}\left(\hat{Q}^{R(A)}\partial_{\bm{r}}\hat{Q}^{R(A)}\right)+i\varepsilon\hat{\tau}_{z}\hat{Q}^{R(A)}-
i\varepsilon'\hat{Q}^{R(A)}\hat{\tau}_{z}+ i\big[\hat{\Delta},\hat{Q}^{R(A)}\big]
-\big[\hat{\Sigma}^{R(A)},\hat{Q}^{R(A)}\big]=0\,.
\end{equation}
The self-energy $\hat{\Sigma}^{R/A}$ captures the effects of pair breaking. For example, in experiments of Ref. \cite{KhrapaiArxiv2}, the leading source of pair breaking seems to be magnetic disorder. Without loss of generality we take 
\begin{equation}\label{eq:Sigma}
\hat{\Sigma}^{R(A)}=\pm\frac{\Gamma_\phi}{2}\hat{\tau}_{z}\,
\end{equation}
with the phenomenologically introduced dephasing rate $\Gamma_\phi=\tau^{-1}_\phi$. Above $T_c$, the superconducting order parameter 
\begin{equation}\label{eq:Delta}
\hat{\Delta}=\left(\begin{array}{cr}0 & \Delta
\\ -\Delta^{*} & 0 \end{array}\right)\,,
\end{equation}
is zero on average $\langle\hat{\Delta}\rangle=0$. However its fluctuations to quadratic order generate corrections to the current in Eq. \eqref{eq:J}. Since current is bilinear in $\hat{Q}$, we have to determine $\hat{Q}^{R/A}$ to linear order in $\hat{\Delta}$. For that purpose we need a particular parameterization for the Keldysh block. We apply the standard decomposition \cite{Kopnin}
\begin{equation}\label{eq:Q-K}
\hat{Q}^{K}_{\varepsilon\varepsilon'}(\bm{r})=\hat{Q}^{R}_{\varepsilon\varepsilon'}(\bm{r})\hat{F}_{\varepsilon'}(\bm{r})-
\hat{F}_{\varepsilon}(\bm{r})\hat{Q}^{A}_{\varepsilon\varepsilon'}(\bm{r})\,,
\end{equation}
where the electronic distribution matrix function is introduced in the form
\begin{equation}\label{n}
\hat{F}_{\varepsilon}(\bm{r})=\left(\begin{array}{cc}
f[\varepsilon-e\phi(\bm{r})] & 0
\\ 0 & f[\varepsilon+e\phi(\bm{r})] \end{array}\right),
\end{equation}
with $\phi(\bm{r})$ being the electric potential. In equilibrium $f_\varepsilon=\tanh(\varepsilon/2T)$.

We proceed to solve Eq. \eqref{eq:Usadel} in perturbation theory over the fluctuating order parameter $\hat{\Delta}(\mathbf{r},t)$. 
To this end, we seek solution in the form (in the Fourier representation)
\begin{equation}\label{eq:Q-RA}
\hat{Q}^{R(A)}_{\varepsilon\varepsilon'}(\bm{r})= \hat{\Lambda}^{R(A)}_{\varepsilon-\varepsilon'}+
\hat{G}^{R(A)}_{\varepsilon\varepsilon'}(\bm{r})\,,\qquad \hat{\Lambda}^{R(A)}_{\varepsilon-\varepsilon'}=\pm2\pi\delta(\varepsilon-\varepsilon')\hat{\tau}_{z}
\end{equation}
where the first term corresponds to the spatially homogeneous normal state ($\Delta=0$), whereas the correction $\hat{G}^{R(A)}$ captures 
inhomogeneous part linear in $\hat{\Delta}_{\varepsilon-\varepsilon'}(\bm{r})$. With the ansatz in the form of Eq.~\eqref{eq:Q-RA} used in Eq.~\eqref{eq:Usadel}, one finds
\begin{equation}\label{eq:G-RA-corr}
D\partial^{2}_{\bm{r}}\hat{G}^{R(A)}_{\varepsilon\varepsilon'}\pm
i(\varepsilon+\varepsilon')\hat{G}^{R(A)}_{\varepsilon\varepsilon'}
-\Gamma_\phi\hat{G}^{R(A)}_{\varepsilon\varepsilon'}=\pm
2i\hat{\Delta}_{\varepsilon-\varepsilon'}\,.
\end{equation}
To arrive at this equation one must respect the nonlinear constraint $\hat{Q}^2=1$ imposed on the $Q$-matrix. This in particular implies that the linear correction  
$\hat{G}^{R(A)}$ has to anticommute with $\hat{\Lambda}$, namely $\big\{\hat{G},\hat{\Lambda}\big\}=0$. This in turn means that $\hat{G}$ has to be off-diagonal. 
Corrections to the diagonal elements of $\hat{G}^{R(A)}$ appear only in the quadratic order in $\hat{\Delta}$. These corrections correspond to the depletion of the density of states 
by superconducting fluctuations. As already mentioned in the introduction, we neglect these terms as they result in the less singular corrections as compared to MT and AL contributions.  

At this point it is convenient to introduce the resolvent for the differential operator on the right-hand-side of Eq. \eqref{eq:G-RA-corr}. 
It defines the retarded/advanced Cooperon propagator that is solution of the following equation
\begin{equation}\label{eq:C}
\left(D\partial^2_{\bm{r}}\pm i\epsilon-\Gamma_\phi\right)C^{R(A)}_{\epsilon}(\bm{r},\bm{r}')=2i\delta(\bm{r}-\bm{r}').
\end{equation}
Knowledge of the Cooperon enables us to write the solution for the Green's function in the compact form 
\begin{equation}\label{eq:G-RA-corr-sol}
\hat{G}^{R(A)}_{\varepsilon\varepsilon'}(\bm{r})=\pm\int\d \bm{r}'
C^{R(A)}_{\varepsilon+\varepsilon'}(\bm{r},\bm{r}')\hat{\Delta}_{\varepsilon-\varepsilon'}(\bm{r}')\,. 
\end{equation}
This step completes the task of finding the $\hat{Q}$-matrix and gives us all the needed ingredients to determine the current density in Eq. \eqref{eq:J}. 

\subsection{Expression for the current}

In order to establish the direct connection to the existing diagrammatic calculations we partially split Eq. \eqref{eq:J} into several district terms 
\begin{subequations}
\begin{eqnarray}
&&\bm{j}=\bm{j}_{\text{N}}+\bm{j}_{\text{AL}}+\bm{j}_{\text{MT}}\,,\\
&&\bm{j}_{\text{N}}=\frac{\pi}{2}e\nu D\,\Tr\left\{\hat{\tau}_{z}\partial_{\bm{r}}\hat{F}_{\varepsilon}\right\}=\sigma_{\text{N}}\bm{E}\,, \label{eq:J-N}\\
&&\bm{j}_{\text{AL}}=\frac{\pi}{2}e\nu D\,\Tr\left\{\hat{\tau}_{z}\left[\hat{G}^{R}_{\varepsilon\varepsilon'}\partial_{\bm{r}}
\hat{G}^{R}_{\varepsilon'\varepsilon}\hat{F}_{\varepsilon}-
\hat{F}_{\varepsilon}\hat{G}^{A}_{\varepsilon\varepsilon'}
\partial_{\bm{r}}\hat{G}^{A}_{\varepsilon'\varepsilon}\right]\right\}\,,\label{eq:J-AL}\\
&&\bm{j}_{\text{MT}}=-\frac{\pi}{2}e\nu D\,\Tr\left\{\hat{\tau}_{z}
\hat{G}^{R}_{\varepsilon\varepsilon'}\partial_{\bm{r}}\hat{F}_{\varepsilon'}\hat{G}^{A}_{\varepsilon'\varepsilon}\right\}\,.\label{eq:J-MT}
\end{eqnarray}
\end{subequations}
Here we introduced the normal state Drude conductivity $\sigma_{\text{N}}=e^2\nu D$ and the electric field $\bm{E}=-\partial\phi/\partial \bm{r}$.  
For compactness, the notation for the global trace $\Tr\{...\}$ incorporates also energy integrations explicit in Eq. \eqref{eq:J} in addition to the matrix trace $\tr\{...\}$. 
As mentioned above, density of stats term $\bm{j}_{\text{DoS}}$ does not appear at the level of our approximation.  It will be shown below that Eq. \eqref{eq:J-AL} and \eqref{eq:J-MT} are in one-to-one correspondence with the respective AL and MT terms. Indeed, when using Eqs. \eqref{eq:Q-RA} and \eqref{eq:G-RA-corr-sol} in Eq. \eqref{eq:J-MT} we have explicitly 
\begin{equation}
\bm{j}_{\text{MT}}=-\frac{\pi}{2}\sigma_{\text{N}}\frac{\partial\phi}{\partial\bm{r}}\iint\frac{\d\varepsilon\d\varepsilon'}{4\pi^2}\iint\d\bm{r}_{1}\d
\bm{r}_2\,C^{R}_{\varepsilon+\varepsilon'}(\bm{r},\bm{r}_1)C^{A}_{\varepsilon+\varepsilon'}(\bm{r},\bm{r}_2)
\tr\left\{\hat{\tau}_{z}\hat{\Delta}_{\varepsilon-\varepsilon'}(\bm{r}_1)
\partial_{\varepsilon'}\hat{F}\hat{\Delta}_{\varepsilon'-\varepsilon}(\bm{r}_2)\right\}\,.
\end{equation}
We change integration variables to the sum and relative energies: $\epsilon=(\varepsilon+\varepsilon')/2$, $\omega=\varepsilon-\varepsilon'$, compute the matrix trace, and introduce the correlation function 
for the average of the product of two order parameter fields $\langle\ldots\rangle$. This gives, at the intermediate step,   
\begin{eqnarray}
&&\bm{j}_{\text{MT}}=-\frac{\pi}{2}\sigma_{\text{N}}\frac{\partial\phi}{\partial
\bm{r}}\iint\frac{\d\epsilon\d\omega}{4\pi^2}\iint\d \bm{r}_{1}\d \bm{r}_2\,
C^{R}_{2\epsilon}(\bm{r},\bm{r}_1)C^{A}_{2\epsilon}(\bm{r},\bm{r}_2)\nonumber\\
&&\times\left[\left\langle\Delta_{\omega}(\bm{r}_1)\Delta^{*}_{-\omega}(\bm{r}_2)\right\rangle
\partial_\epsilon f(\epsilon-\omega/2+e\phi)+
\left\langle\Delta^{*}_{\omega}(\bm{r}_1)\Delta_{-\omega}(\bm{r}_2)\right\rangle
\partial_\epsilon f(\epsilon-\omega/2-e\phi)
\right]\,.
\end{eqnarray}
The two terms in the square brackets can be combined together after an interchange of integration variables 
$\omega\to-\omega$, $\epsilon\to-\epsilon$ simultaneously with an exchange of spatial coordinates $\bm{r}_1\rightleftarrows
\bm{r}_2$. Lastly, integrating by parts one arrives at 
\begin{equation}\label{eq:J-MT-1}
\bm{j}_{\text{MT}}=\pi\sigma_{\text{N}}\frac{\partial\phi}{\partial\mathbf{r}}\iint\frac{\d\epsilon\d\omega}{4\pi^2}\iint\d \bm{r}_{1}\d \bm{r}_2\,\frac{\partial}{\partial\epsilon}
\left[C^{R}_{2\epsilon}(\bm{r},\bm{r}_1)C^{A}_{2\epsilon}(\bm{r},\bm{r}_2)\right]
\left\langle\Delta_{\omega}(\bm{r}_1)\Delta^{*}_{-\omega}(\bm{r}_2)\right\rangle
f(\epsilon-\omega/2+e\phi)\,.
\end{equation}

In complete analogy, we can derive the corresponding expression for $\bm{j}_{\text{AL}}$ in Eq. \eqref{eq:J-AL}. Making use of the commutation relation
$[\hat{\tau}_{z},\hat{f}_\varepsilon]=0$, one finds from Eq. \eqref{eq:J-AL}
\begin{eqnarray}
\bm{j}_{\text{AL}}=\frac{\pi}{2}e\nu
D\iint\frac{\d\varepsilon\d\varepsilon'}{4\pi^2}\iint\d \bm{r}_1\d
\bm{r}_2\,\left[C^{R}_{\varepsilon+\varepsilon'}(\bm{r},\bm{r}_1)\partial_{\bm{r}}C^{R}_{\varepsilon+\varepsilon'}(\bm{r},\bm{r}_2)-C^{A}_{\varepsilon+\varepsilon'}(\bm{r},\bm{r}_1)\partial
_{\bm{r}}C^{A}_{\varepsilon+\varepsilon'}(\bm{r},\bm{r}_2)\right]\nonumber\\
\times\tr\left\{\hat{\tau}_z\hat{\Delta}_{\varepsilon-\varepsilon'}(\bm{r}_1)
\hat{\Delta}_{\varepsilon'-\varepsilon}(\bm{r}_2)\hat{F}_\varepsilon\right\}\,.
\end{eqnarray}
Using $\epsilon,\omega$ energy variables as above, taking the matrix trace, and introducing the average for the order parameter fluctuations, one finds 
\begin{eqnarray}\label{eq:J-AL-1}
\bm{j}_{\text{AL}}=\frac{\pi}{2}e\nu
D\iint\frac{\d\epsilon\d\omega}{4\pi^2}\iint\d \bm{r}_1\d\bm{r}_2\,
\left[C^{R}_{2\epsilon}(\bm{r},\bm{r}_1)\partial_{\bm{r}}C^{R}_{2\epsilon}(\bm{r},\bm{r}_2)-C^{A}_{2\epsilon}(\bm{r},\bm{r}_1)\partial
_{\bm{r}}C^{A}_{2\epsilon}(\bm{r},\bm{r}_2)\right]\nonumber\\
\times\left[\left\langle\Delta^{*}_{\omega}(\bm{r}_1)\Delta_{-\omega}(\bm{r}_2)\right\rangle
f(\epsilon+\omega/2+e\phi)-
\left\langle\Delta_{\omega}(\bm{r}_1)\Delta^{*}_{-\omega}(\bm{r}_2)\right\rangle
f(\epsilon+\omega/2-e\phi) \right]\,.
\end{eqnarray}

To complete the calculation of currents, we need to perform statistical averaging over fluctuations. To accomplish this, we recall that the dynamics of the order parameter is governed by the time-dependent Ginzburg-Landau (TDGL) equation \cite{GE68}
\begin{equation}\label{eq:TDGL}
\left(D\partial^2_{\bm{r}}+i\omega-2ie\phi(\bm{r})-\tau^{-1}_{\text{GL}}-\Gamma_\phi\right)\Delta_{\omega}(\bm{r})=\Upsilon_\omega(\bm{r})\,,
\end{equation}
there the stochastic field $\Upsilon_\omega(\bm{r})$ captures random Langevin forces \cite{TDGL}. In an arbitrary nonequilibrium situation, $\Upsilon_\omega(\bm{r})$ is a complicated functional of the electronic distribution function $f$. It is a challenging task to determine its precise form in general. To make further progress we will rely on the following experimentally motivated approximation. If the relaxation time $\tau_{\text{R}}$ of the electronic system is long as compared to the fluctuation relaxation time $\tau_{\text{GL}}$ of $\Delta$, fluctuations quickly adjust to the local electronic disequilibrium. Therefore, fluctuations of current (and conductivity) are primarily dominated by fluctuations of the distribution function occurring at long time scales rather than by the dynamics of $\Delta$ at shorter times. This can be justified in the limit not too close to $T_c$ where it is sufficient to use the equilibrium form of the Langevin flux averaged with the local temperature $T(\bm{r})$        
\begin{equation}\label{eq:Delta-corr}
\left\langle\Upsilon_{\omega}(\bm{r})\Upsilon^{*}_{-\omega}(\bm{r}')\right\rangle=
\frac{16T^2}{\pi\nu}\delta(\bm{r}-\bm{r}')\,.
\end{equation}
We will further justify this approximation by showing that temperature fluctuations lead to a smaller effect on the current noise. To find $\Delta_\omega(\bm{r})$ from Eq. \eqref{eq:TDGL}, we introduce the resolvent of the corresponding operator, the fluctuation propagator, that satisfies    
\begin{equation}\label{eq:L}
\left(D\partial^2_{\bm{r}}\pm
i\omega-2ie\phi(\bm{r})-\tau^{-1}_{\text{GL}}-\Gamma_\phi\right)L^{R(A)}_{\omega}(\bm{r},\bm{r}')=\delta(\bm{r}-\bm{r}').
\end{equation}
Then the solution to Eq. \eqref{eq:TDGL} can be written as an integral  
\begin{equation}\label{eq:Delta-L}
\Delta_{\omega}(\bm{r})=\int\d \bm{r}'
L^{R}_{\omega}(\bm{r},\bm{r}')\Upsilon_{\omega}(\bm{r}')\,.
\end{equation}
With this input we can perform averages in Eqs. \eqref{eq:J-MT-1} and \eqref{eq:J-AL-1}. For the MT-process contribution to the current we determine  
\begin{equation}\label{eq:J-MT-2}
\bm{j}_{\text{MT}}=16e^2DT^2\frac{\partial\phi}{\partial\bm{r}}
\iint\frac{\d\epsilon\d\omega}{4\pi^2}\,\iiint\d \bm{r}_1\d \bm{r}_2\d\bm{r}_3
\partial_\epsilon [C^{R}_{2\epsilon}(\bm{r},\bm{r}_1)C^{A}_{2\epsilon}(\bm{r},\bm{r}_2)]
L^{R}_{\omega}(\bm{r}_1,\bm{r}_3)L^{A}_{\omega}(\bm{r}_2,\bm{r}_3)
f(\epsilon-\omega/2+e\phi)\,.
\end{equation}
In parallel, for the AL-process we derive    
\begin{eqnarray}\label{eq:J-AL-2}
\bm{j}_{\text{AL}}=8eDT^2\iint\frac{\d\epsilon\d\omega}{4\pi^2}\iiint\d \bm{r}_1\d
\bm{r}_2\d \bm{r}_3\,\left[C^{R}_{2\epsilon}(\bm{r},\bm{r}_1)\partial_
{\bm{r}}C^{R}_{2\epsilon}(\bm{r},\bm{r}_2)-C^{A}_{2\epsilon}(\bm{r},\bm{r}_2)
\partial_{\mathbf{r}}C^{A}_{2\epsilon}(\bm{r},\bm{r}_2)\right]\nonumber\\
\times L^{R}_{\omega}(\bm{r}_{1},\bm{r}_3)L^{A}_{\omega}(\bm{r}_{2},\bm{r}_3)
\left[f(\epsilon+\omega/2+e\phi)-f(\epsilon+\omega/2-e\phi)\right]\,.
\end{eqnarray}
The above expressions for currents are complimentary to the well-known diagrammatic counterparts but apply to a broader class of problems including nonequilibrium fluctuations. 
In particular, they can be used to describe fluctuations in the spatially inhomogeneous superconducting systems and mesoscopic contacts. 

\subsection{Fluctuation conductivity}

It is useful to perform a consistency check. In the linear response analysis to a spatially uniform field all propagators depend only on the coordinate differences, namely, 
$\big\{C^{R(A)}_{\epsilon}(\bm{r},\bm{r}'),L^{R(A)}_{\omega}(\bm{r},\bm{r}')\big\}\to\big\{
C^{R(A)}_{\epsilon}(\bm{r}-\bm{r}'),L^{R(A)}_{\omega}(\bm{r}-\bm{r}')\big\}$. Therefore, it is convenient to transform real space integrals into the momentum representation:
\begin{equation}
\iiint\d \bm{r}_1\d \bm{r}_2\d \bm{r}_3
C^{R}_{2\epsilon}(\bm{r}-\bm{r}_1)C^{A}_{2\epsilon}(\bm{r}-\bm{r}_2)
L^{R}_{\omega}(\bm{r}_1-\bm{r}_3)L^{A}_{\omega}(\bm{r}_2-\bm{r}_3)=
\sum_{q}|C^{R}_{2\epsilon}(q)|^2|L^{R}_{\omega}(q)|^2\,.
\end{equation}
In accordance with our convention, the Fourier transform of Eqs.~\eqref{eq:C} and \eqref{eq:L} gives
\begin{equation}
|C^R_{\epsilon}(q)|^2=\frac{4}{(Dq^2+\Gamma_\phi)^2+\epsilon^2}\,, \quad
|L^R_{\omega}(q)|^2=\frac{1}{(Dq^2+\tau^{-1}_{GL}+\Gamma_\phi)^2+\omega^2}\,.
\end{equation}
which leads to the following expression for the MT-conductivity:
\begin{equation}\label{eq:sigma-MT}
\sigma_{\text{MT}}=\frac{16e^2 DT^2}{\pi^2}\sum_{q}\iint\d\epsilon\d\omega
\frac{\partial_{\epsilon}f_\epsilon}
{\big[(Dq^2+\Gamma_\phi)^2+(2\epsilon+\omega)^2\big]
\big[(Dq^2+\tau^{-1}_{\text{GL}}+\Gamma_\phi)^2+\omega^2\big]}\,.
\end{equation}
A quick inspection suggests that the characteristic energies of bosonic modes are determined by the inverse GL time $\{Dq^2,\omega\}\sim \tgl^{-1}\sim T-T_c$. This simply follows from the fluctuation propagator. 
In contrast, fermionic energies entering Cooperon and the distribution function are of the order of temperature $\epsilon\sim T$. Owing to this energy scale separation, and being interested only in the most singular piece of the integral in $T-T_c\ll T_c$, it is sufficient to approximate $\partial_\epsilon f_\epsilon\to1/(2T)$. Energy integrations then separate into two simple Lorentzian functions so that     
\begin{equation}\label{eq:sigma-MT1}
\sigma_{\text{MT}}=4e^2DT\sum_q\frac{1}{(Dq^2+\tau^{-1}_{\text{GL}})(Dq^2+\Gamma_\phi)}.
\end{equation}    
In this expression, we absorbed the dephasing factor $\Gamma_\phi$ into the shift of $T_c$, namely $\tau^{-1}_{\text{GL}}+\Gamma_\phi\to\tau^{-1}_{\text{GL}}$ with $T_c\to T_{c0}-\pi\Gamma_\phi/8$. For the two-dimensional case the final momentum integral reproduces the celebrated result \cite{MakiPTP68,ThompsonPRB70}
\begin{equation}\label{eq:sigma-MT2}
\sigma_{\text{MT}}=\frac{e^2}{8}\frac{\ln(\epsilon_{\text{T}}/\gamma_\phi)}{\epsilon_{\text{T}}-\gamma_\phi},\quad \epsilon_{\text{T}}=\frac{T-T_c}{T_c}, \quad \gamma_\phi=\frac{\pi\Gamma_\phi}{8T_c}. 
\end{equation}
An analogous computation for the AL-term shows that it is smaller by a factor of two in the prefactor and does not contain logarithm. In what follows, we will work in the limit of weak dephasing where $\ln(\epsilon_{\text{T}}/\gamma_\phi)>1$. Since we are interested in the current-current correlation function we focus on MT contribution which is dominant over the AL term. This is also supported in applications to 2D films by microscopic calculations of fluctuation conductivity in mesoscopic superconductor-normal metal devices, where it was shown that in the planar NSN contacts the MT contribution dominates over the whole temperature range near $T_c$ \cite{VolkovPRB98}. Since noise power scales as the square of the current, the combination of the numerical factor and logarithm gives a large parameter in favor of MT term. 

\subsection{Excess noise in fluctuation region}

The obtained expressions for the currents in Eqs. \eqref{eq:J-MT-2} and \eqref{eq:J-AL-2} suggest that fluctuations in the electronic distribution $\delta f$ generate the respective fluctuations in conductivities. This motivates consideration of the following noise function of conductivity fluctuations:
\begin{equation}\label{eq:S-sigma}
\delta S_\sigma(\Omega)=\frac{1}{\mathcal{V}}\int \langle\delta\sigma(\bm{r},t)\delta\sigma(\bm{r}',t')\rangle e^{i\Omega(t-t')}\d(t-t')\d\bm{r}', 
\end{equation}
where $\mathcal{V}$ is the sample volume. For the MT process specifically we have 
\begin{equation}\label{eq:sigma-f}
\delta\sigma(\bm{r},t)=\int\d\epsilon\chi(\epsilon)\delta f(\epsilon,\bm{r},t),\quad \chi(\epsilon)=-\frac{4e^2DT^2}{\pi^2}\sum_q\int\d\omega\partial_\epsilon|C^R_{2\epsilon+\omega}(q)|^2|L^R_{\omega}(q)|^2.
\end{equation}
It is worthwhile to notice that in contrast to conductivity fluctuations in normal metals, the susceptibility $\chi(\epsilon)$ is strongly energy dependent. Therefore, fluctuations at different energies become correlated in the Cooper channel. The correlation function for the distribution function fluctuations can be determined in the Boltzmann-Langevin theory. It requires finding the solution to the kinetic equation for the leading mechanism of electron relaxation. For our purposes, it suffices to take the relaxation time approximation, in which the correlation function takes a standard form
\begin{equation}\label{eq:f-f}
\langle\delta f(\varepsilon,\bm{r},t)\delta f(\varepsilon',\bm{r}',t')\rangle_\Omega=\frac{4\tau_{\text{R}}/\nu}{1+(\Omega\tau_{\text{R}})^2}\delta(\varepsilon-\varepsilon')
\delta(\bm{r}-\bm{r}')f_\varepsilon[1-f_\varepsilon]. 
\end{equation}
With the help of Eqs. \eqref{eq:sigma-f} and \eqref{eq:f-f}, we find the following expression for the intrinsic conductivity noise in the vicinity of superconducting phase transition
\begin{equation}
\delta S_\sigma(\Omega)=\left(\frac{4e^2 DT^2}{\pi^2}\right)^2\frac{(4\tau_{\text{R}}/\nu\mathcal{V})}{1+(\Omega\tau_{\text{R}})^2}
\sum_{qq'}
\iint\d\epsilon\d\omega\d\omega'
f_\epsilon[1-f_\epsilon]
\partial_\epsilon|C^R_{2\epsilon+\omega}(q)|^2
\partial_\epsilon|C^R_{2\epsilon+\omega'}(q')|^2
|L^R_{\omega}(q)|^2|L^R_{\omega'}(q')|^2
\end{equation} 
We are interested in the most singular term in powers of $T-T_c$. This can be extracted with the same sequence of steps as in the conductivity calculation. Taking advantage of the energy scale separation between bosonic and fermionic frequencies it is sufficient to replace $f_\epsilon[1-f_\epsilon]\to1$. 
This approximation enables one to complete $\epsilon$-integration in the closed form 
\begin{equation}
\int^{+\infty}_{-\infty}\d\epsilon \partial_\epsilon|C^R_{2\epsilon+\omega}(q)|^2
\partial_\epsilon|C^R_{2\epsilon+\omega'}(q')|^2=64\pi\frac{\mathcal{D}_\phi+\mathcal{D}'_\phi}{\mathcal{D}_\phi\mathcal{D}'_\phi}
\frac{3(\omega-\omega')^2-(\mathcal{D}_\phi+\mathcal{D}'_\phi)^2}{[(\mathcal{D}_\phi+\mathcal{D}'_\phi)^2+(\omega-\omega')^2]^3}.
\end{equation} 
Here we introduced the shorthand notations $\mathcal{D}_\phi=Dq^2+\Gamma_\phi$ and $\mathcal{D}'_\phi=Dq'^2+\Gamma_\phi$. 
This leaves us with the subsequent double-$\omega$ integration which also admits a fully analytical result that can be found with the help of the identity 
\begin{equation}
\iint^{+\infty}_{-\infty}\frac{[3(\omega-\omega')^2-(\mathcal{D}_\phi+\mathcal{D}'_\phi)^2]\d\omega\d\omega'}
{[\mathcal{D}^2_{\text{GL}}+\omega^2][\mathcal{D}'^2_{\text{GL}}+\omega'^2][(\mathcal{D}_\phi+\mathcal{D}'_\phi)^2+(\omega-\omega')^2]^3}=\frac{\pi^2}
{8\mathcal{D}_{\text{GL}}\mathcal{D}'_{\text{GL}}(\mathcal{D}_\phi+\mathcal{D}'_\phi)(\Gamma_{\text{GL}}-\Gamma_\phi)^3}.
\end{equation}
In analogy to the above expression we introduced an additional shorthand notation $\mathcal{D}_{\text{GL}}=Dq^2+\Gamma_{\text{GL}}$. 
When combining the last two integrals we notice that the factor $\mathcal{D}_\phi+\mathcal{D}'_\phi$ mixing momenta cancels out. 
This significantly simplifies the remaining momentum integrations that factorize into the product 
\begin{equation}
\sum_{qq'}[\mathcal{D}_\phi\mathcal{D}'_\phi\mathcal{D}_{\text{GL}}\mathcal{D}'_{\text{GL}}]^{-1}=\left(\sum_q[\mathcal{D}_\phi\mathcal{D}_{\text{GL}}]^{-1}\right)^2
\end{equation}
Finally, noticing that the particular combination $\sum_q[\mathcal{D}_\phi\mathcal{D}_{\text{GL}}]^{-1}$ appears in the definition of the MT conductivity Eq. \eqref{eq:sigma-MT1}, 
we arrive at our main result for the conductivity noise   
\begin{equation}
\delta S_\sigma(\Omega)=\frac{32\sigma^2_{\text{MT}}\tau_{\text{R}}}{\pi\nu\mathcal{V}T[1+(\Omega\tau_{\text{R}})^2]}\left(\frac{T}{\Gamma_{\text{GL}}-\Gamma_\phi}\right)^3.
\end{equation}
For the superconducting film we convert this power spectrum of conductivity fluctuations into the corresponding spectral density of correlated excess current noise $\delta S_{\text{corr}}$. 
For that purpose we use the identities $I=LE\sigma_{\text{MT}}$ and $\delta S_{\text{corr}}=L^2E^2\delta S_\sigma$, where $E$ is the electric field and $L$ is the side length of a square sample. 
Then using the expression for the MT conductivity from Eq. \eqref{eq:sigma-MT2}, we arrive at the main result of this section
\begin{equation}\label{eq:S-I-sigma}
\delta S_{\text{corr}}(\Omega)=\frac{32}{\pi g}\left(\frac{E_{\text{Th}}}{T_c}\right)(I^2\tau_{\text{R}})\frac{(T_c\tgl)^3}{1+(\Omega\tau_{\text{R}})^2},
\end{equation}
where $g$ is the dimensionless film conductance which was defined earlier in the introduction. In addition, for simplicity, we replaced $T\to T_c$ everywhere except in $\tau_{\text{GL}}$ and took $\Gamma_{\text{GL}}>\Gamma_\phi$ consistently with earlier assumptions. This is the origin of the result Eq. \eqref{eq:S-corr} quoted in the introduction. 


\section{Discussion}\label{sec:discussion}

\subsection{Comparison and estimates}

The theoretical analysis presented in the preceding sections demonstrates sensitivity of the excess noise to three time scales: $\tgl$, $\tau_\phi$ and $\tau_{\text{R}}$.
The analysis of experimental data in Refs. \cite{KhrapaiArxiv1,KhrapaiArxiv2} suggests that the relaxation is dominated by the electron-phonon interaction, 
whereas dephasing is primarily mediated by magnetic disorder. The expression for the excess noise from fluctuations of the MT-conductivity Eq. \eqref{eq:S-I-sigma} 
was derived based on the assumption of fluctuations of the electronic distribution function. However, given the extreme sensitivity of the fluctuation conductivity to $\tgl$
temperature fluctuations may also trigger strong current noise. We therefore estimate zero-frequency excess noise from the temperature fluctuations 
\begin{equation}
\delta S_{\text{temp}}=E^2L^2\left(\frac{\d\sigma_{\text{MT}}}{\d T}\right)^2\langle\delta T^2\rangle\tau_{\text{R}}=E^2L^2\sigma^2_{\text{MT}}\frac{T^2_c\tau_{\text{R}}}{C_{\text{el}}L^2d(T-T_c)^2}=
\frac{64}{\pi^2}I^2\tau_{\text{R}}\frac{(T_c\tgl)^2}{C_{\text{el}}L^2d}.
\end{equation}
Here we used the standard textbook expression for the mean square of temperature fluctuations that is expressed in terms of the electronic heat capacity per unit volume $C_{\text{el}}$ \cite{LL-Vol5}. 
We also retained only the most singular term in the temperature derivative of the MT conductivity. Using the heat capacity of the Fermi gas $C_{\text{el}}=\pi^2\nu T_c/3$ evaluated at $T=T_c$ and re-expressing density of states in the Sommerfeld coefficient through the sheet conductance in the normal state $\nu=\sigma^{\square}_{\text{N}}/(e^2dD)$ we find equivalently 
\begin{equation}\label{eq:S-I-T}
\delta S_{\text{temp}}=\frac{192}{\pi^4g}\left(\frac{E_{\text{Th}}}{T_c}\right)(I^2\tau_{\text{R}})(T_c\tgl)^2.
\end{equation}
This estimate can be readily compared to Eq. \eqref{eq:S-I-sigma}. The ratio of the two contribution to the excess noise is $\delta S_{\text{temp}}/\delta S_{\text{corr}}\sim \epsilon_{\text{T}}<1$, 
which implies that the correlated conductivity fluctuations mechanism of noise dominates over the temperature fluctuation mechanism. We remind the reader that the applicability range for this estimate requires 
$\tgl<\{\tau_\phi,\tau_{\text{R}}\}$. The two main factors limiting $\tau_\phi$ are magnetic disorder and intrinsic effect. The former factor results in $\tau_\phi\sim 5$ ps for 5 nm thick TiN films in the experiments of Ref. \cite{KhrapaiArxiv1}. The latter factor can be estimated from the nonlinear fluctuation theory \cite{Reizer92,LO2001,AL2010} to be $\tau^{-1}_{\phi}\sim T_c\sqrt{Gi}$, where $Gi=1/23g$ is the Ginzburg number. It results in $\tau_\phi\sim50$ ps for TiN films with $g\sim40$. Therefore, at the threshold of theory applicability $\tgl\sim\tau_\phi$ the ratio of two noise terms is  $\delta S_{\text{temp}}/\delta S_{\text{corr}}\sim 10^{-2}$ for the intrinsic dephasing mechanism, and  $\delta S_{\text{temp}}/\delta S_{\text{corr}}\sim 10^{-1}$ for the dephasing by magnetic impurities. This indicates that the calculated correlated conductivity noise can be a more important fluctuation mechanism in a broad range of parameters near $T_c$.

\begin{figure}
\centering
\includegraphics[scale=0.55]{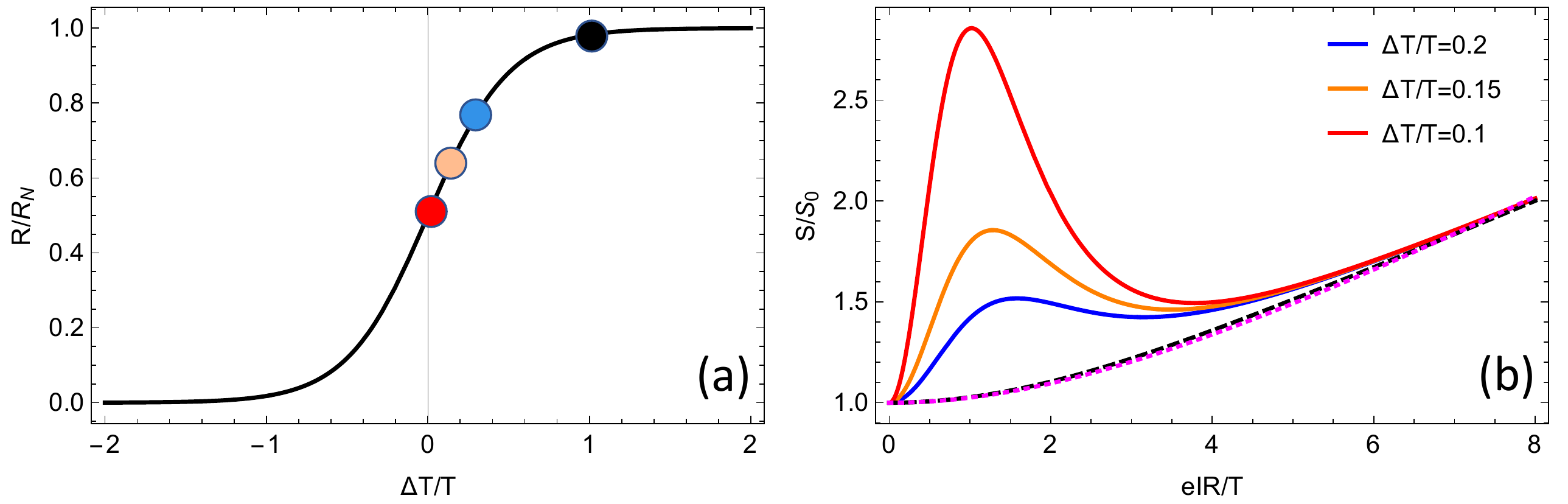}
\caption{Schematic representation of the superconducting transition shown by the resistive curve on panel (a) 
and the concurrent current noise on panel (b) presented at different temperature points along the resistive line. 
$R(T)$ is normalized to the value in the normal state $R_{\text{N}}$, whereas noise is normalized to its equilibrium 
Johnson-Nyquist value $S_0$ at $T_c$. The apparent excess noise peak manifests close to the critical temperature. }
\label{Fig-Noise}
\end{figure}

\subsection{Summary and outlook}

In this work we presented a comprehensive overview of the fluctuation-driven intrinsic mechanisms of excess noise in superconducting films.  These mechanisms include direct contribution of
superconducting fluctuations near $T_c$ to transport properties [Eqs. \eqref{eq:S-reg} and \eqref{eq:S-anom}], electron heating [Eq. \eqref{eq:S-heat}], correlated conductivity fluctuations [Eq. \eqref{eq:S-I-sigma}], and temperature fluctuations [Eq. \eqref{eq:S-I-T}]. In order to highlight the main features of the obtained results, on Fig. \ref{Fig-Noise} we schematically reproduce experimentally observed noise across the superconducting transition. The panel (a) shows the resistive transition as a function of temperature. The colored circles on the $R(T)$ line indicate specific temperatures at which noise is then measured. Panel (b) shows the respective noise curves as a function of bias current in the sample.  The dashed lines represent the reference to the spectral noise in the normal state that display crossover from the thermal noise to the shot noise. In fact, there are two lines in that plot for the regimes of cold and hot electrons per Eqs. \eqref{eq:noise-N1} and \eqref{eq:noise-N2}, which are almost indistinguishable for the chosen parameters. When current noise is measured at temperatures that are progressively closer to $T_c$, a large peak of excess noise develops. The analytical expressions presented in this work provide perturbative corrections to the noise power $\propto I^2$ and thus capture its uprise on top of the base line. The full crossover was calculated earlier for zero-dimensional NSN quantum dot devices. It takes the form \cite{DBAL}
\begin{equation}
\frac{\delta S_{\text{exc}}}{S_0}\simeq\left(\frac{E_{\text{Th}}}{\Delta T}\right)^2\frac{Y^2}{(a+bY^2)^3},\qquad Y=\frac{eV}{\sqrt{T_c\Delta T}},\qquad \Delta T=T-T_c, 
\end{equation}  
and successfully describes both uprise $\propto Y^2$ and fall off $\propto 1/Y^4$ of the noise signal. In the formula $a$ and $b$ are model specific constant parameters. 
In our calculations we were unable to derive the whole crossover curve, but we expect a qualitatively similar behavior of the noise function.  
In the zero-dimensional case, the Usadel equation is just a matrix equation, rather than a differential equation for the matrix field as in the two-dimensional geometry [Eq. \eqref{eq:Usadel}], so further analytical progress was possible. Extensions of the theory in 2D case to capture the whole curve is thus still an open problem. Another outstanding problem concerns generalization of these results to the regime below $T_c$. The existing theories address the resistive curve $R(T)$, see for example an overview given in Ref. \cite{KonigPRB15}, however, no such description was developed for the noise. Below $T_c$ one naturally expects phase slips to play an important role. For the Poissonian process of independent phase slips, based on dimensional analysis, one can anticipate shot noise $S\propto \Phi_0V$, where $\Phi_0$ is the quantum flux quantum. Further experimental insights into the mechanisms of noise in superconducting devices can be provided by measurements of switching current distributions \cite{BezryadinPRL13}. We anticipate that future research will bring new results in this important area of physics.    

\section{Acknowledgments}

We are grateful to Vadim Khrapai for the earlier collaboration \cite{KhrapaiArxiv1} that initiated and stimulated this work. 
We thank Robert McDermott for bringing Ref. \cite{IrwingHilton} to our attention, and acknowledge Felix Jaeckel and Dan
McCammon for discussion on the topic of noise mechanisms in the context of superconducting transition-edge sensors.
This work was supported by the U.S. Department of Energy (DOE), Office of Science, Basic Energy Sciences (BES) under Award No. DE-SC0020313.
A.L. is grateful to the Max Planck Institute for Solid State Research for hospitality, where this work was performed in part, 
and to the Alexander von Humboldt Foundation for the financial support of the visit.

\section*{References}

\bibliography{biblio}

\end{document}